\documentstyle[prb,aps,manuscript,epsfig]{revtex}  
\begin{document}                
\title{Ultrasound Investigations of\\ Orbital Quadrupolar Ordering in UPd$_3$}
\author{N. Lingg, D. Maurer, and V. M\"uller}
\address{Institut f\"ur Physik, Universit\"at Augsburg, Universit\"atsstra\"se 1, 86159 Augsburg, Germany}

\author{K. A. McEwen}
\address{Department of Physics and Astronomy, University College London, Gower
Street,\\
London WC1E 6BT, United Kingdom}

\maketitle
\begin{abstract}                
For a high-quality single crystal of UPd$_3$ we present the relevant elastic constants and
ultrasonic attenuation data. In addition to the magnetic phase transition
at $T_2=4.4 \pm 0.1$K and the quadrupolar transition at $T_1\sim6.8$K, we find orbital ordering at
$T_0=7.6 \pm 0.1$K concomitant with a symmetry change from hexagonal to orthorhombic.
A striking feature is the splitting of the phase transition at $T_1$ into a second-order transition
at $T_{+1}=6.9 \pm 0.05$K and a first-order transition at $T_{-1}=6.7 \pm 0.05$K. For the four
phase transitions, the quadrupolar order parameters and the respective symmetry changes are
specified.

\vspace{10pt}

\noindent PACS Numbers: 71.45.-d, 62.20.Dc, 71.70.Ms
\end{abstract}

\vspace{30pt}
The actinide intermetallic nonheavy fermion system UPd$_3$ and the heavy fermion superconductor UPt$_3$
are hexagonal compounds with space group $P6_3/mmc$, exhibiting magnetic phase transitions at $T_2=4.4$K and $\sim5$K,
respectively. This finding is very surprising, because in both compounds the ordered magnetic moment is extremely small
($\mu \approx 10^{-2}\mu_B$). Furthermore, the $5f$ electrons of UPt$_3$ are found to be predominantly itinerant,
whereas UPd$_3$ belongs to the rather few metallic materials where well-localized $5f$ electrons and ionic long-range
quadrupolar ordering at $T_1=6.8$K has been identified. The possibility that quadrupolar order occurs at $T_1$
was first recognized in specific heat~\cite{1} and thermal expansion~\cite{2} measurements, and the strong quadrupole-quadrupole
interaction between the U ions was confirmed for the first time in sound velocity experiments~\cite{3},
where also the relevant quadrupole components contributing to the magnetoelastic coupling were specified.
The ultimate proof of the quadrupolar nature of the phase transition, however, was given in neutron scattering
experiments~\cite{4}, where the long-range lattice distortions which always accompany~\cite{6} a quadrupolar phase transition
were determined.

The main reason for the interest in the quadrupolar phase transition of UPd$_3$ is that the knowledge of the ordering
components of the quadrupole tensor not only provides detailed information about the long-range ordering of
the $5f\,^2$ electronic charge distribution of the uranium U$^{4+}$ ions (ground state $^3$H$_4$), but also of
long-range spatial correlations of the orbital degrees of freedom. This is so, because the average values of
the components $Q_{ij}$ of the electric quadrupole tensor {\bf Q} are not only a measure of deviations of the
localized $5f$ uranium electrons from a spherical charge distribution but, according to the identity
$Q_{ij} \propto \{3(J_iJ_j + J_jJ_i)/2 + J(J+1)\}$, are also a measure of correlations between
the components $J_i$ and $J_j$ of the total angular momentum {\bf J}. We mention that long-range quadrupolar
(i.e., orbital) ordering is also of current interest in the context of the physics behind the colossal magnetoresistance
in the manganates~\cite{6} or the spin-Peierls and metal-insulator transitions~\cite{6} in the vanadates.

In order to deduce from the lattice distortions the ordering quadrupole components $Q_{ij}$ (i.e., the order parameters),
the complete strain tensor {\boldmath$\varepsilon$} or, more strictly speaking, the symmetry adapted strains $\varepsilon_{\Gamma}$
must be known because the latter are proportional~\cite{5} to the average value of the symmetry adapted components of the
quadrupole tensor. In neutron diffraction experiments, however, the full strain tensor is difficult to deduce from
the available Bragg peaks. This is one of the reasons why details concerning the various phase transitions and the
proper order parameters in UPd$_3$ are still under discussion. Also, it is still an open question to what extent
the inter-ion quadrupole-quadrupole interaction is responsible for the three reported phase transitions found at
$T_2=4.4$K, $T_1=6.8$K and $T_0=7.8$K~\cite{7}.

It is well known that ultrasound is well suited~\cite{8} to answer these questions because the coupling of the acoustic
strain {\boldmath$\varepsilon$} to the ionic quadrupole moment tensor {\bf Q} is, in general, large and the coefficients of the elastic
stiffness tensor {\bf C} depend in a characteristic manner~\cite{5,8} (see Table \ref{table1}) on the components of the so-called
quadrupolar strain susceptibility $\chi^{(Q)}$. The latter is defined as $\delta \mbox{\bf Q} = \chi^{(Q)}\mbox{\bf V}$
and therefore is a measure of the response $\delta \mbox{\bf Q}$ of the quadrupole moment tensor {\bf Q} to the
strain-induced electric field gradient tensor {\bf V} at the uranium sites. The components $V_{ij}$ are related
to the elastic strain tensor {\boldmath$\varepsilon$} via $V_{ij} = \sum S_{ijkl} \varepsilon_{kl}$, where the fourth-rank
tensor {\bf S} is the so-called {\it field gradient elastic strain} tensor which may be determined~\cite{9} in nuclear
acoustic resonance experiments: this should not be confused with the elastic compliance tensor. The quadrupolar
contribution $C_{\Gamma}^{(Q)}$ to a symmetry elastic constant $C_{\Gamma}$ is given
by~\cite{8,10}
\begin{displaymath}
C_{\Gamma}^{(Q)} = -N_V \frac{(S_{\Gamma}^{(0)})^2 \cdot \chi_{\Gamma}^{(Q)}}{1-g'_{\Gamma} \chi_{\Gamma}^{(Q)}}
\end{displaymath}
where $N_V$ is the number density of quadrupoles, $g'_{\Gamma}$ the two ion quadrupole-quadrupole coupling constant
and (depending on the irreducible representation $\Gamma$), $S_{\Gamma}^{(0)}$ is a linear combination of the S-tensor
components for $g'_{\Gamma}=0$ (i.e., in the absence of the inter-ion quadrupole-quadrupole interaction).
In terms of fluctuations [i.e., of the variance $(\Delta Q_{\Gamma})^2$] of the symmetry adapted quadrupole tensor
components $Q_{\Gamma}$, the quadrupolar strain susceptibility $\chi_\Gamma^{(Q)}$ may be written in the thermostatic
(zero-frequency) limit as $\chi_\Gamma^{(Q)} \propto (\Delta Q_{\Gamma})^2 /$ $(k_BT)$, where, in general, five relevant
quadrupole tensor components exist which, in the symmetry adapted form, we denote by
$Q_{zz}$, $Q_{x^2-y^2} = Q_{xx}-Q_{yy}$, $Q_{xy}$, $Q_{yz}$, and $Q_{zx}$. If $Q_{\Gamma}$ is the order parameter,
then a pronounced decrease of the related elastic constants $C_{\Gamma}$ (see Table \ref{table1}) should be observed at a
second-order quadrupolar phase transition, because in that case, the fluctuations of $Q_{\Gamma}$ (and accordingly the strain
susceptibility $\chi_{\Gamma}^{(Q)}$) should become large. Provided that at a quadrupolar phase transition the formation
of domains or its influence on the elastic behavior can be minimized such as to play only a minor role, then it
becomes immediately evident that ultra-sound should be well suited to study long-range quadrupolar (i.e., orbital)
ordering.

The efficiency of ultrasound is demonstrated in Figs.~1(a) and 2(a) for longitudinal acoustic waves propagating along
the c axis of a rectangular single crystal of UPd$_3$ (with the linear dimensions 3.6 x 5.2 x 7.3 mm$^3$).
As can be clearly seen, both the attenuation (phonon loss rate) and the sound velocity of the $C_{33}$ mode display
three distinct phase transitions at $T_2=4.4 \pm 0.1$K, $T_1 \approx 6.8$K and $T_0=7.6 \pm 0.1$K, whereas the
specific heat~\cite{11} and magnetic susceptibility~\cite{12} of samples of the same batch exhibit pronounced peaks only
at $T_1$, and $T_1$ and $T_2$, respectively. Previous sound velocity~\cite{3,13} and neutron scattering experiments~\cite{14}
revealed antiferroquadrupolar ordering at $T_1$ (Refs.~13 and 14) on the quasi-cubic uranium sites and it was thought that
the concomitant structural transformation is from hexagonal to trigonal~\cite{7,14} or to orthorhombic~\cite{3} or monoclinic~\cite{3}.
At first sight, it is surprising that this transition dominates the specific heat but (compared to the magnitude of
the ultrasound absorption peaks observed at $T_2$ and $T_0$) contributes only little to the phonon-loss rate.
According to Table \ref{table1} and to Refs.~5 and 15, on the other hand, $\langle  Q_{zz}\rangle$ is not the proper order parameter of
a quadrupolar phase transition concomitant with a structural transformation from hexagonal to trigonal
(monoclinic or orthorhombic). We therefore do not expect that the phonon-loss rate of the $C_{33}$ mode will be
modified essentially in the vicinity of $T_1$ which is in agreement with our experimental findings (see Fig.~1).
The absence of a huge absorption peak at $T_1$ (see Fig.~1) of both the $C_{33}$ and the $C_{44}$ mode further
indicates that at $T_1$ the contribution of incoherent strains (originating from domains or domain walls) may be
considered as negligibly small. This suggests that the ordered phase is the quadrupolar triple-{\bf q} state~\cite{14},
because the triple-{\bf q} state avoids the formation of stochastically distributed domains. The development of the
triple-{\bf q} state is associated with modulations of the electronic charge distribution. These modulations are
a superposition of three plane waves with wave vectors {\bf q}$_1$, {\bf q}$_2$ and {\bf q}$_3$, where the wave
vectors {\bf q}$_2$, and {\bf q}$_3$ are obtained from {\bf q}$_1$ by rotations of $\pm 2\pi / 3$. The trigonal
triple-{\bf q} phase (space group $P\overline{3}m1$) was discovered~\cite{14} in neutron diffraction experiments and,
within experimental uncertainty, the respective phase transition at $T_1$ was found to be continuous~\cite{14}.
As we will see below, however, this is in contradiction to our findings.

We have mentioned already that the average value $\langle  Q_{zz}\rangle$ cannot serve as a primary order parameter because
it is already nonzero above $T_1$ (and $T_0$). The formation of the triple-{\bf q} phase may, nevertheless,
modify the $C_{33}$ mode because the triple-{\bf q} state alters~\cite{14} the $Q_{zz}$ moments of the cubic site ions.
Surprisingly (see Fig.~2) the $C_{33}$ mode does not soften at $T_1$ upon cooling (as expected for a continuous
phase transition), but shows a steplike increase which suggests that the transition at $T_1$ is most likely of
first order. Even more surprising is the finding [see inset of Fig.~1(a)] that at around $T_1$ not one but two clearly
distinguishable absorption peaks (and accordingly two phase transitions) appear at $T_{-1} = 6.7$K and $T_{+1} = 6.9$K,
where the stiffness coefficients $C_{33}$ and $C_{44}$ turn out to be hysteretic in the vicinity of $T_{-1}$
(see Fig.~3) but not at $T_{+1}$. In addition, [see Fig.~1(b)] we also have observed at $T_{-1}$ a pronounced
hysteresis in the attenuation of the $C_{44}$ mode. The quadrupolar transition into the trigonal triple-{\bf q}
state must therefore be of first order, whereas the transition at $T_{+1}$ is most likely of second order.
We would like to emphasize that, to the best of our knowledge,
for UPd$_3$ neither ultrasound absorption measurements nor the existence of the
two neighboring phase transitions at $T_1$ have been previously reported. Compared to other experimental techniques,
it is furthermore worthwhile to note (see Figs.~1 and 2) that at zero magnetic field the phase transition at $T_0=7.6$K
is most apparent in ultrasound.

So far, we have confined ourselves mainly to the $C_{33}$ mode which only reflects the temperature dependence
of the secondary order parameter $\langle  Q_{zz}\rangle$. For hexagonal symmetry (see Table \ref{table1}) a softening of the $C_{44}$ mode
is attributed to an ordering of the quadrupole components $\langle  Q_{yz}\rangle$ and/or $\langle  Q_{zx}\rangle$, whereas the order
parameters related to  $C_{66} = (C_{11} - C_{12})/2$ (or to $C_{11}$ and $C_{22}$) are the degenerate
quadrupoles $\langle  Q_{xy}\rangle$ (or $\langle  Q_{x^2-y^2}\rangle$). We now consider the temperature dependence of $C_{44}$ and $C_{66}$.
As can be seen in Fig.~2, with decreasing temperature, $C_{44}$ exhibits a small dip at $T_{-1}$, followed by a
steep decrease at the magnetic transition temperature $T_2=4.4$K which is accompanied by a very sharp peak
(see Fig.~1) in the phonon loss rate. However, the most pronounced softening of the $C_{66}$ mode appears at
$T_0=7.6$K, not at $T_1$ as reported for a sample of much lower crystal quality and size by other authors~\cite{3}.
The softening of the $C_{44}$ and $C_{66}$ modes, which starts already far above the transition temperatures
$T_2$ and $T_0$, is very strong and it has been shown~\cite{13} that below 150K the temperature dependence of the
elastic constants is essentially due to the crystal-field splitting at the quasicubic uranium sites.
We therefore conclude that not only at $T_1$, but also at $T_0$ and at the magnetic transition temperature
$T_2$, the driving force of the respective phase transitions is of quadrupolar origin. Hence, the quadrupolar
order parameter of the magnetic phase ($T < T_2$) is most likely (see below) a linear combination~\cite{15} of
$\langle  Q_{yz}\rangle$ and $\langle  Q_{x^2-y^2}\rangle$, whilst at $T_0$ the relevant order parameters are the degenerate quadrupoles
$\langle  Q_{xy}\rangle$ and $\langle  Q_{x^2-y^2}\rangle$. Here, {\it degenerate} means that the response of the respective quadrupoles is
specified by the same strain susceptibility.

The degenerate order parameters $\langle  Q_{x^2-y^2}\rangle$ and $\langle  Q_{xy}\rangle$ which, according
to Table \ref{table1}, will modify
$C_{11}$, $C_{22}$ and $C_{66} = (C_{11}-C_{12})/2$, are associated with a transition from hexagonal to
orthorhombic~\cite{15,16} or to monoclinic~\cite{16}, whereas the order parameters $\langle  Q_{yz}\rangle$ and $\langle  Q_{zx}\rangle$ will change
$C_{44}$ and $C_{55}$ and entail a transition from a hexagonal to either a monoclinic or triclinic phase~\cite{15,16}.
Here, the formation of a monoclinic or triclinic phase depends on whether or not both order parameters are present
simultaneously. Since no anomaly is seen at $T_0=7.6$K in   the elastic modulus $C_{44}$, whereas the
$C_{11}$ and $C_{22}$ modes (which are not shown here) and the $C_{66}$ mode soften significantly, it follows that
at $T_0$ we pass with decreasing temperature from the hexagonal to an orthorhombic or monoclinic phase.
When analyzing the phase transitions below $T_0$, we therefore have to take into account that for
$T_{+1} < T < T_0$, the symmetry is orthorhombic or monoclinic but not hexagonal (which in the literature
was taken for granted until recently). The monoclinic phase, however, may be excluded because in that case
the only possible second-order transition at $T_{+1}$ would be~\cite{15} from a monoclinic to a triclinic
phase, which is in contrast to our experimental findings (see Fig.~3), because such a transition should alter
$C_{44}$ significantly. At $T_0$ the symmetry of UPd$_3$, therefore, passes at decreasing temperature
from hexagonal to orthorhombic and the only possible-second order transition~\cite{15,16} at $T_{+1}$ is
from the orthorhombic to a monoclinic state. At $T_{+1}$ we therefore have to consider~\cite{15,16}
$\langle  Q_{yz}\rangle$, $\langle  Q_{zx}\rangle$, and $\langle  Q_{xy}\rangle$ as possible order parameters, where $\langle  Q_{yz}\rangle$ and $\langle  Q_{zx}\rangle$ may
be excluded~\cite{15,16} because $C_{44}$ (see Fig.3) and $C_{55}$ do not alter significantly at $T_{+1}$.
Concerning $\langle  Q_{xy}\rangle$, however, which should modify~\cite{15} $C_{66}$, we cannot prove experimentally whether
or not $\langle  Q_{xy}\rangle$ becomes an order parameter at $T_{+1}$, because in the temperature range between 6.6K and
7.8K, the attenuation of the $C_{66}$ mode becomes too large. We note that in agreement with Ref.~\cite{15}
and our experimental findings, the phase transition at $T_{-1}$ (i.e., from the monoclinic to the trigonal
triple-{\bf q} state) is of first order. The softening of the $C_{44}$ mode at $T_2=4.4$K further
indicates~\cite{15} that at decreasing temperature, UPd$_3$ transforms at $T_2$ from the trigonal triple-{\bf q}
state into a monoclinic (magnetic) state, where the transition is expected~\cite{15} to be of first order.
Within experimental resolution, however, the $C_{44}$ mode does not show a signature of a first-order
transition which indicates that at $T_2$ it is possibly not enough to consider the magnetoelastic coupling
only. According to Ref.~15 the transition from the triple-{\bf q} into the monoclinic phase should be
accompanied by the loss of the ternary axis and the order parameter should be a linear combination of
$\langle  Q_{yz}\rangle$ and $\langle  Q_{x^2-y^2}\rangle$. It is there-fore expected that both $C_{44}$ and $C_{66}$ will soften at
$T_2$, which is confirmed by our experiments (see Fig.~2). In the immediate vicinity of $T_2$, however, the
attenuation of the $C_{66}$ mode becomes too large to enable a more detailed analysis of this phase transition.

Based upon the crystal-field split ($J = 4$) multiplets determined by Buyers {\it et al.},\cite{17} the temperature
dependence of $C_{44}$ has been evaluated previously in Ref.~13 yielding the quadrupolar coupling
constants $N_V(S_{44}^{(0)})^2/C_{44}^{(0)} $ $= 0.6$K and $g'_{44} = -6$K. Since the ordered magnetic moments
in UPd$_3$ are so extremely small~\cite{4} and the quadrupolar coupling constants are so large, the pronounced softening
of the $C_{44}$ mode indicates that the quadrupole-quadrupole interaction is the most probable driving
mechanism of the magnetic phase transition at $T_2$. Owing to the negative sign of the quadrupole coupling
constant $g'_{44}$ we further expect~\cite{10,13} antiferroquadrupolar ordering at $T_2$ and the formation of a
very complex antiferromagnetic state, which is currently under investigation and will be the subject of a
forthcoming paper. Summarizing our results, we have shown that ultrasound is a powerful tool for
investigating orbital ordering and that the large variety of interesting phase transitions in UPd$_3$ is
of quadrupolar origin. We hope that our findings may also offer some new ideas in the attempt of a better
understanding of the formation of the heavy fermion state and the antiferromagnetic transition in UPt$_3$.

This work was supported by the Deutsche Forschungsgemeinschaft (Grant No.~MU 696/ 2-2). We acknowledge
many helpful and clarifying discussions with K.-H.~H\"ock.

\newpage

\begin{figure}[htb]
   \begin{center}
      \epsfig{file=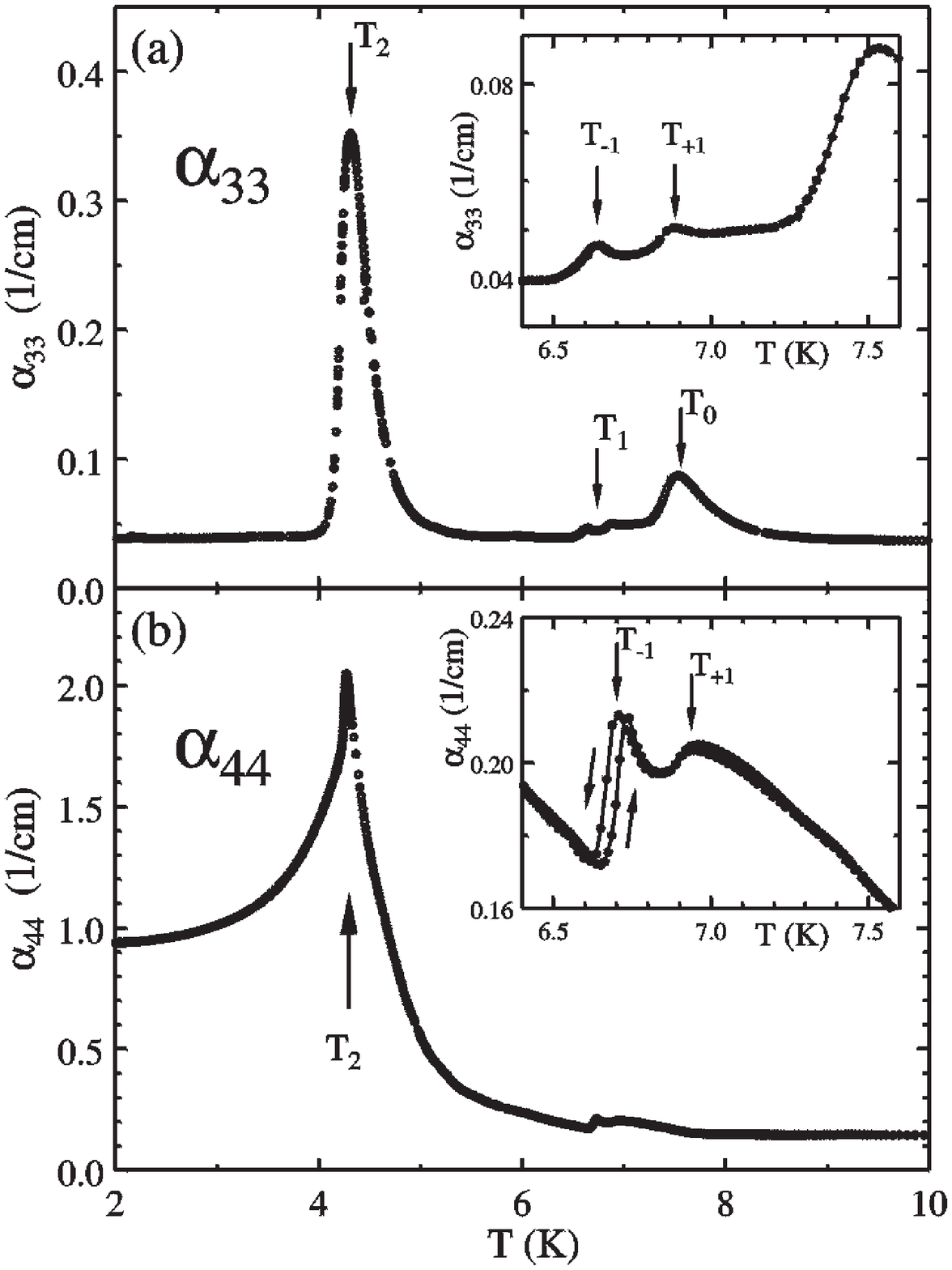,width=12cm}
   \end{center}
   \caption{\small Temperature dependence of the $\sim20$MHz longitudinal (a) and transverse (b) attenuation coefficients
of the $C_{33}$ and $C_{44}$ elastic modes.}
   \label{eps1}
\end{figure}

\newpage

\begin{figure}[htb]
   \begin{center}
      \epsfig{file=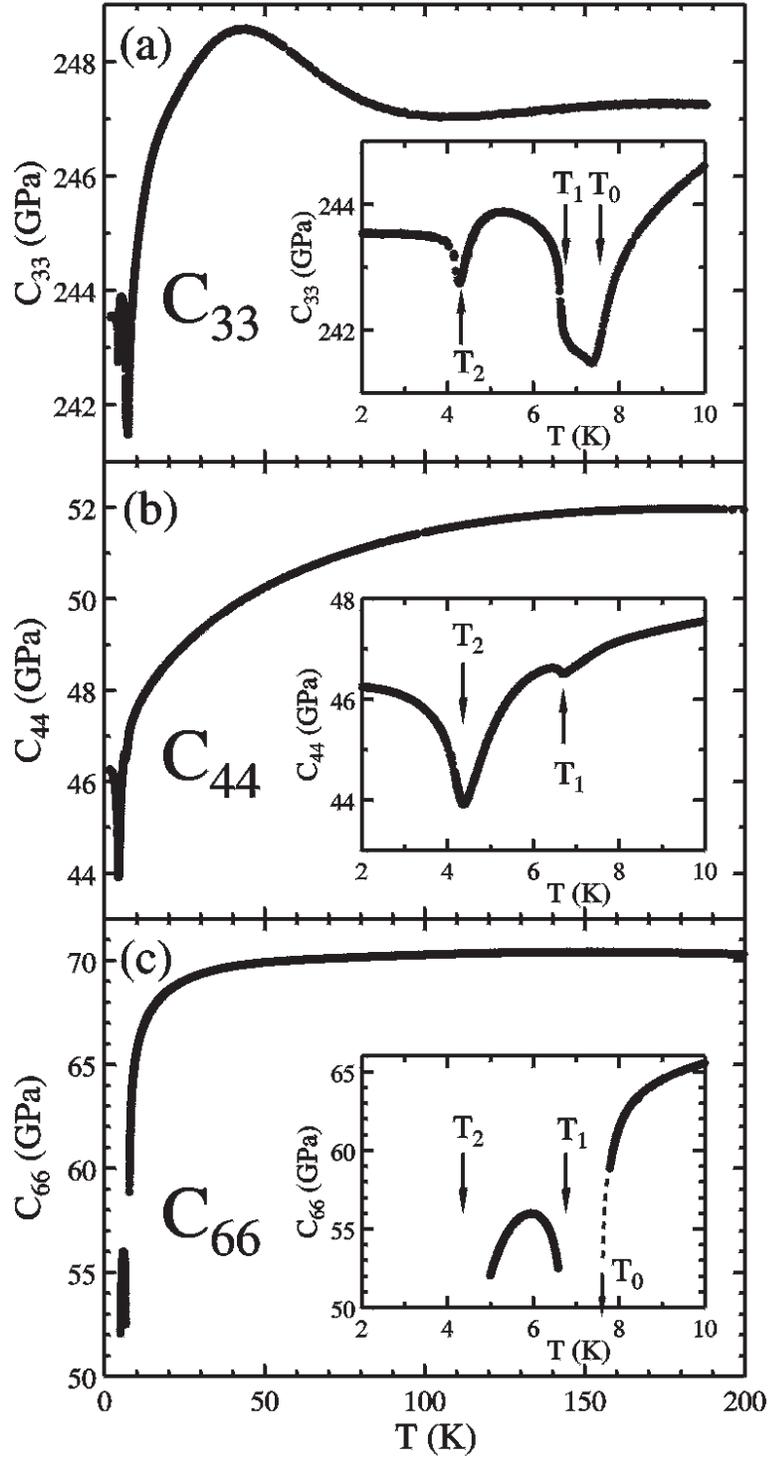,width=12cm}
   \end{center}
   \caption{\small Temperature dependence of the elastic constants $C_{33}$, $C_{44}$, and $C_{66}$ at about 20MHz.}
   \label{eps2}
\end{figure}

\newpage

\begin{figure}[htb]
   \begin{center}
      \epsfig{file=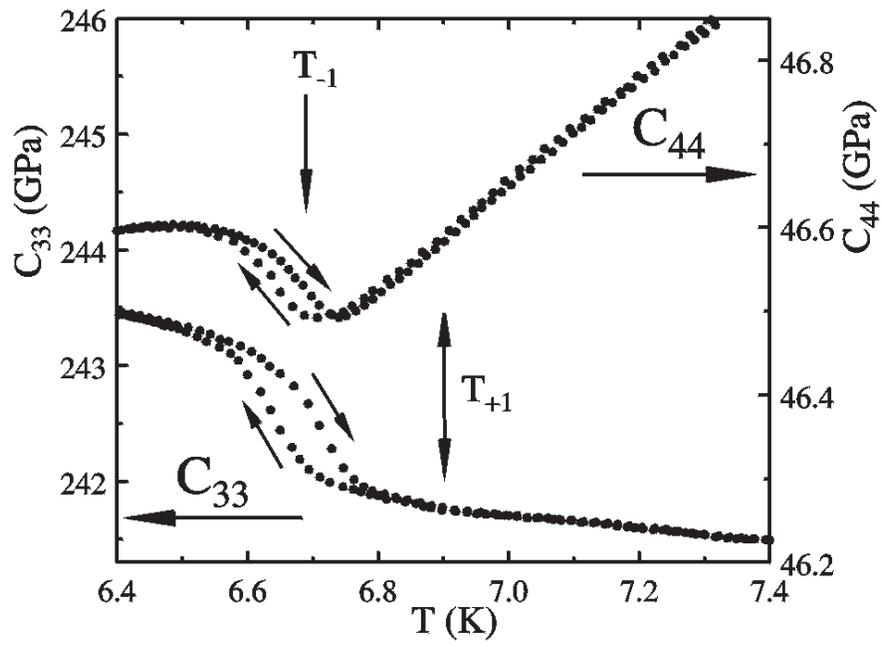,width=12cm}
   \end{center}
   \caption{\small Hysteretic behavior of the $C_{33}$ and $C_{44}$ elastic mode at about 22MHz.}
   \label{eps3}
\end{figure}

 \begin{table}
 \caption{\small Quadrupolar contribution $C_{\alpha \beta}^{(Q)}$ to the elastic constants in Voigt notation,
the strain-induced quadrupole interaction Hamiltonian H$_Q$ and the relevant strain susceptibilities
$\chi_{\Gamma}^{(Q)}$ for hexagonal crystal symmetry and different acoustic strain modes $\bf \varepsilon_{ij}$.
The indices x, y, and z refer to the a-, b-, and c-axis of the crystal.}

\vspace{10pt}

\begin{tabular}{c @{\qquad\quad} c @{\qquad\quad} c @{\qquad\quad} c}
$ \varepsilon_{ij}$ & $ C_{\alpha \beta}^{(Q)}$ & $ H_Q$ & $ \chi_{\Gamma}$\\
\tableline
$\varepsilon_{xx}$ & $C_{11}^{(Q)}$ & $\{-(S_{11}+S_{12})Q_{zz}+(S_{11}-S_{12})Q_{x^2-y^2}\}\varepsilon_{xx}$ & $\chi_{zz}^{(Q)}$, $\chi_{x^2-y^2}^{(Q)}$  \\
$\varepsilon_{yy}$ & $C_{22}^{(Q)}$ & $\{-(S_{11}+S_{12})Q_{zz}-(S_{11}-S_{12})Q_{x^2-y^2}\}\varepsilon_{yy}$ & $\chi_{zz}^{(Q)}$, $\chi_{x^2-y^2}^{(Q)}$  \\
$\varepsilon_{zz}$ & $C_{33}^{(Q)}$ & $S_{33}Q_{zz}\varepsilon_{zz}$ & $\chi_{zz}^{(Q)}$ \\
$\varepsilon_{yz}$ & $C_{44}^{(Q)}$ & $S_{44}Q_{yz}\varepsilon_{yz}$ & $\chi_{yz}^{(Q)}$ \\
$\varepsilon_{zx}$ & $C_{55}^{(Q)}$ & $S_{44}Q_{zx}\varepsilon_{zx}$ & $\chi_{zx}^{(Q)} = \chi_{yz}^{(Q)}$ \\
$\varepsilon_{xy}$ & $C_{66}^{(Q)}$ & $(S_{11}-S_{12})Q_{xy}\varepsilon_{xy}$ & $\chi_{xy}^{(Q)} = \chi_{x^2-y^2}^{(Q)}$
 \end{tabular}
 \label{table1}
 \end{table}

\end{document}